\documentclass[pdflatex,sn-basic]{sn-jnl}

\usepackage{graphicx}%
\usepackage{multirow}%
\usepackage{booktabs}

\usepackage{amsmath,amssymb,amsfonts}%
\usepackage{amsthm}%
\usepackage{mathrsfs}%
\usepackage[title]{appendix}%
\usepackage{xcolor}%
\usepackage{textcomp}%
\usepackage{manyfoot}%
\usepackage{booktabs}%
\usepackage{algorithm}%
\usepackage{algorithmicx}%
\usepackage{algpseudocode}%
\usepackage{listings}%
\usepackage{tabulary}
\usepackage{tabularx}
\usepackage{diagbox}
\usepackage{slashbox}
\usepackage{float}

\theoremstyle{thmstyleone}%
\theoremstyle{thmstyletwo}%

\theoremstyle{thmstylethree}%

\begin{document}

\title[]{The Flawed Foundations of Fair Machine Learning}


\author*[1,2]{\fnm{Robert Lee} \sur{Poe}}\email{roberlee.poe@santannapisa.it}

\author[2,3]{\fnm{Soumia Zohra} \sur{El Mestari}}\email{soumia.elmestari@uni.lu}

\affil*[1]{\orgdiv{LIDER Laboratory}, \orgname{Sant'Anna School of Advanced Studies}, \orgaddress{\street{Via Santa Cecilia 3}, \city{Pisa}, \postcode{56127}, \country{Italy}}}

\affil[2]{\orgdiv{Interdisciplinary Center for Security, Reliability and Trust}, \orgname{University of Luxembourg}, \orgaddress{\street{6 avenue de la Fonte}, \city{Esch-sur-Alzette}, \postcode{L-4364}, \country{Luxembourg}}}


\abstract{The definition and implementation of fairness in automated decisions has been extensively studied by the research community. Yet, there hides fallacious reasoning, misleading assertions, and questionable practices at the foundations of the current fair machine learning paradigm. Those flaws are the result of a failure to understand that the trade-off between statistically accurate outcomes and group similar outcomes exists as independent, external constraint rather than as a subjective manifestation as has been commonly argued. First, we explain that there is only one conception of fairness present in the fair machine learning literature: group similarity of outcomes based on a sensitive attribute where the similarity benefits an underprivileged group. Second, we show that there is, in fact, a trade-off between statistically accurate outcomes and group similar outcomes in any data setting where group disparities exist, and that the trade-off presents an existential threat to the equitable, fair machine learning approach. Third, we introduce a proof-of-concept evaluation to aid researchers and designers in understanding the relationship between statistically accurate outcomes and group similar outcomes. Finally, suggestions for future work aimed at data scientists, legal scholars, and data ethicists that utilize the conceptual and experimental framework described throughout this article are provided.\footnote{This article is a preprint submitted to the Minds and Machines Special Issue on the (Un)fairness of AI on May 31st, 2023.}}




\maketitle

\section{Introduction}\label{sec1}

Automated decision-making systems are increasingly being used to render high-impact decisions regarding human beings. All the while, notorious accounts of algorithmic discrimination and algorithmic unfairness have been reported by news outlets over the past decade. Due to the many concerns about the potential societal impacts of machine learning, governments are beginning to put forward policy positions and draft regulations. In the AI Bill of Rights, the White House states that automated decisions should be designed and deployed to achieve equitable outcomes. In Europe, the AI Act states that automated decisions should not perpetuate historic patterns of discrimination or create new forms of disparate impact. Right now, policy-makers and regulators are relying heavily on the fair machine learning community to present solutions. However, a unipolar conception of fairness is being represented and advocated for by the fair machine learning community, which does not reflect the same breadth of opinion that exists in wider society.\footnote{Problems related to the fair distribution of finite resources are continually addressed in politics, law, and culture with varying views about how best to address those problems.}

To a limited extent, researchers in the field have understood that they had not happened upon an empty field (of research) but instead a garden that has been fostered, cared for, and in some cases ignored for a very long time. Perspectives from many domains have been incorporated into the literature: legal doctrines like disparate impact \citep{barocas_big_2016}, fair distribution philosophies dealing with egalitarianism and merit \citep{arif_khan_towards_2022}, socio-technical critiques of technological solutionism \citep{cooper_emergent_2021,selbst_fairness_2019}, and concepts from feminist communications and data science like the myth of objectivity and meritocracy \citep{gebru_race_2020,dignazio_6_2020}. Due to the multidisciplinary nature of the field, a sufficient framework for understanding the technical limitations and implications of machine learning combined with the goals and implications of ``fairness'' in between and within related disciplines has yet to be fully realized. This is partly due to the fact that most articles published from the machine learning community are difficult for those in the humanities or legal communities to understand and vice versa. Thus researchers at the crossroads of fairness, discrimination, and machine learning are to some degree speaking past one another. Throughout the entire article, each audience is kept in mind and essential terminology is defined.

In section \ref{fairnessmetrics}, we argue that the common categorization of fairness in the literature misleading implies that there are three competing conceptions of fairness, when in fact there is one: fairness defined as the similarity in outcomes between groups when that similarity acts in the benefit of underprivileged groups. In section \ref{tradeoff}, we argue that the trade-off between statistically accurate outcomes and group similar outcomes is an independent external constraint on fair machine learning, rather than a mere framing problem as is commonly argued. And, we address some misconceptions about the implications of that trade-off found in the literature. In section \ref{framework1}, we carefully address a specific example, the highly influential view proposed by Friedler et al. in \citep{friedler_impossibility_2016,friedler_impossibility_2021}. Finally, in section \ref{framework2}, we will introduce a proof-of-concept evaluation to aid researchers and designers in understanding the relationship between statistically accurate outcomes and group similar outcomes.

\section{Fair Machine Learning Means Group Similarity in Outcomes}\label{fairnessmetrics}

In the fairness literature, metrics are commonly split into three categories: group fairness, causal fairness, and individual fairness. This categorization misleading implies that there are three distinguishable conceptions of fairness at play in the fair machine learning literature, when in reality there is but one conception of fairness technically defined in metrics: fairness defined as group similarity in outcomes based on a given sensitive attribute. The mantra of group fairness is that groups should be treated equally or at least similarly. The mantra of group fairness is, however, misleading because "fairness" defined as group similarity ensures the \emph{similarity of outcomes}, not solely the \emph{similarity of treatment}. 

Causal fairness shares the same conception of fairness as group fairness and is only different in so far as a different set of techniques are used to achieve that goal \citep{kilbertus_avoiding_2017, kusner_counterfactual_2017, prost_simpsons_2022,castelnovo_clarification_2022}. The same is true of individual fairness, even though individual fairness has incessantly been offered as an opposing conception of fairness to group fairness definitions with its own mantra of similar individuals should be treated similarly. The maxim of similar treatment that individual fairness embodies is an Aristotelian principle of consistency \citep{binns_apparent_2020}. The individual fairness definition states that there should be consistency between the relevant features of two different persons and their respective outcomes in comparison to one another. More specifically, the similarity between the features of two individuals (measured as a distance) should be preserved between their respective labels.\footnote{Labels or target variables refer to the variable to be predicted by the machine learning algorithm.}

Note that the principle of consistency, defined as distance between spaces, \emph{could} be used to detect whether there exists inconsistencies between the relevant features and ground truth of the sample, as well as when there exists inconsistencies between the sample and the outcomes. For instance, the inconsistencies could be seen as an indicator of unreliable data collection processes where data was incorrectly reported, or that the data sample is missing a set of uncollected features that could explain the current inconsistency. We emphasize, however, that the individual fairness metric itself is not concerned with determining the representativeness of the sample nor with determining how well the outcomes generalize to a target population.

Individual fairness defines \emph{fairness} as a comparison of geometric distances. Once distance is defined, individuals can be compared and inconsistencies (unfairness) can be rectified. However, the distance must be defined, and defining a distance presupposes prior knowledge about ``fairness." In other words, the principle of consistency is \emph{empty} \citep{schauer_treating_2018}, and so requires a substantive notion of fairness to define what makes similar cases similar (i.e. the distance). Thus, there is a circularity in the proposition that individual fairness is a definition of fairness. It may be that the principle of consistency is a necessary requirement for fairness to be achieved, but consistency or similarity alone is not sufficient to constitute an independent notion of fairness \citep{fleisher_whats_2021}. Some advocates of the individual fairness approach argue that substantive notions of fairness need be defined by domain experts \citep{dwork_fairness_2012,friedler_impossibility_2021}, others argue that the distance can be learned \citep{mukherjee_two_2020}, while still others argue that the group fairness metrics should fill the void \citep{fleisher_whats_2021}. In any case, individual fairness should be understood as a tool to implement fairness once defined, rather than as a conception of fairness in and of itself.

Thus, there is but one conception of fairness \emph{technically} defined in fairness metrics: group similarity in outcomes based on a given sensitive attribute. To build machine learning models that produce outcomes that are group similar, the first step is to define a measure or metric that reflects a notion of acceptable group dissimilarity. There exists a ``zoo'' of these metrics \citep[p.2]{castelnovo_clarification_2022} that define the acceptability of group dissimilarity differently using notions of statistical independence, sufficiency, and separation. A number of surveys and reviews on the taxonomy of metrics and interventions have been published \citep{mehrabi_survey_2021, pessach_review_2022, castelnovo_clarification_2022, carey_statistical_2023}.

Once a metric of acceptable group dissimilarity is chosen, one of the three following strategies can be adopted: (1) pre-processing the input data to remove, alter, or curate the underlying data that lead to group dissimilarities \citep{feldman_certifying_2015,zhang_achieving_2017,hajian_methodology_2013}, (2) in-processing where the model is constrained to produce group similar outcomes by modifying the learning algorithm's objective functions \citep{berk_convex_2017,zafar_fairness_2017};
and/or (3) post-processing the output of the model, rather than changing anything about the sample or hypothesis assumptions, by using an algorithm based on a function that detects potential group dissimilarities and adjusts the labels accordingly \citep{kim_multiaccuracy_2019}. 

As many have observed, requiring similarity of outcomes can either act in the benefit or detriment of underprivileged groups depending on the decision context \citep{selbst_fairness_2019,cooper_emergent_2021,carey_statistical_2023}\textemdash where privileged and underprivileged groups are identified by a given legal or moral code. To achieve an equitable outcome, rather than a merely equal or similar one, researchers argue that the choice of whether or not to use fair machine learning or of which metric defining the acceptability of group dissimilarity is to be used, must be context specific so that similarity in outcomes only act in the detriment of privileged groups and in the favor of underprivileged groups. Justifications for such a definition of fairness in machine learning are inspired by (1) legal doctrines like disparate impact and affirmative action in the United States or indirect discrimination and positive action in the European Union \citep{barocas_big_2016,thomas_algorithmic_2021,wachter_why_2021}, (2) critical race and gender studies which find disparities that disfavor underprivileged groups to constitute \emph{de facto} discrimination \citep{gebru_race_2020,dignazio_6_2020}, (3) or distribution philosophies based on moral arguments for the acceptance of equity-based systems and the rejection of merit-based systems \citep{arif_khan_towards_2022} (or a combination of these justifications \citep{carey_statistical_2023}). In the end, fairness in machine learning generated decisions is defined as \emph{similarity in outcomes between groups when that similarity acts in the benefit of underprivileged groups to the detriment of privileged groups.}

\section{The Trade-Off Between Statistically Accurate Outcomes and Group Similar Outcomes}\label{tradeoff}

Statistically accurate outcomes are the result of a robust model trained on a representative sample. Group dissimilarity in outcomes can be the result of (1) an unrepresentative sample and/or non-generalizable assumptions (inaccuracies), (2) group dissimilarities existing in the target population that are reflected in a representative sample and carried into the outcomes by a robust model (accuracies), or (3) unrepresentative sampling and/or hypothesis formulations that add \emph{additional} group dissimilarity beyond the group dissimilarity already present in the target population (both). The goal of traditional machine learning is to produce statistically accurate outcomes. The traditional machine learning approach encompasses a number of techniques and practices to remove inaccuracies including those that would produce more group dissimilarity than that which is present in a target population. A target population can have imbalanced group sizes, contain anomalies like the Simpson's paradox, have sub-group validation problems, and so on.\footnote{A sub group validity problem occurs when a particular observable characteristic is valid for some groups but not for others.} In other words, differences between groups in the target population can result in group dissimilar outcomes. Wrongheadedly, many fall into the trap of thinking that dissimilarities between groups in outcomes \emph{must} be the result of inaccuracies. Statistically inaccurate outcomes can certainly either \emph{exaggerate} or \emph{underestimate} the disparities which exist in a target population, but the solution to statistically inaccurate outcomes is to create a more representative sample and/or robust model.

Researchers should confront the fact that the goal of fair machine learning \emph{is not} to increase the representativeness of a data sample or the robustness of a model. In fact the goal is the opposite. Decisions made on a representative sample have the potential to reflect the target population in the model outcomes, and those outcomes would have the same disparities between groups that exist in the target population. In other words, if the relevant (for the task) base-rate parameters of the target population result in a demographic parity of 0.4, that demographic parity would be the limit of fairness in a model that generalizes to the target population. The realization that there may be differences between groups in the world is a painful one. However, to reject that there exists differences between groups is to reject the existence of ``unfairness.'' In other words, if there were no differences between groups in the target population, there would be no differences in outcomes for the equitable, fair machine learning approach to address. 

To illustrate how difficult it has been for the fair machine learning community to internalize this point, reflect for a moment on the difference in the use of the word ``bias" between traditional machine learning and fair machine learning. Traditionally, bias is defined as a deviation from the true value of a parameter or variable. In fair machine learning, bias is defined as a deviation from group similarity \citep{pessach_review_2022,mehrabi_survey_2021,chen_why_2018,castelnovo_clarification_2022}. When the true value of a parameter leads to group dissimilarity in outcomes, the \emph{true value} is dubbed biased. When the word ``bias" in the sense of deviation from group similarity (normative) is used to reject a true value of a parameter in statistics, the is-ought fallacy is often committed. The is-ought fallacy occurs when one reaches a normative conclusion based solely on descriptive (factual) premises. In this case, by rejecting the true statistical value based on normative concerns, the descriptive aspects of statistics are being mixed with the prescriptive aspects of morality. This terminological mix-up often results in the fallacy of equivocation, which occurs when a keyword of an argument is used with more than one meaning, leading to misinterpretation. Disguising normative judgements in the language of statistics only creates confusion.

While philosophers might best understand the thrust of the argument through examples of the is-ought fallacy, jurists might best understand by comparing the Separation Thesis found in legal positivism to the differentiation made here \citep{hart_positivism_1957}. The separation thesis insists on the separation between (1) what the law \emph{is} and (2) what the law \emph{ought} to be. Here, there is a separation between what \emph{is} (accuracy) and what \emph{ought} to be (fair). In other words, the traditional machine learning approach has been a descriptive effort and the fair machine learning approach has been a prescriptive effort.

Data scientists can only ever follow best practices (under epistemological limitations) when sampling from the target population, including ensuring a sample with a relatively small standard deviation, adequate size, random sampling, and so on. Following best practices will, admittedly, never constitute a proof of representativeness. However, the lack of proof does not mean that a data scientist is unfamiliar with the target population or that they are unable to estimate group disparity. The reader may be tempted, in response to the arguments made throughout this section, to doubt whether awareness of group dissimilarity in a target population is practically possible. Note though, if awareness of group dissimilarity is not practically possible, then there would be no evidence of group dissimilarity\textemdash which many researchers in the field use as a proxy for \emph{de facto} discrimination\textemdash and so no \emph{justification} for the disparate impact legal doctrine.

The trade-off between statistically accurate outcomes and group similar outcomes is obvious. The logical conclusion of the trade-off is also obvious: where there exists the greatest need for the equitable, fair machine learning approach (i.e. data settings that contain large group disparities), machine learning itself is most useless.\footnote{Where "need" is defined by a normative judgement.} As the connection between the model outcomes and the target population becomes more tenuous to become less dissimilar amongst sub-groups, the use of statistical learning becomes harder to justify. In other words, the more the outcome is already known (manually coded), the less need there is for a data driven approach. A script or quota could fulfill the same purpose. Thus, the trade-off presents an existential threat to the field.

The response to the threat of the trade-off between statistically accurate outcomes and group similar outcomes has been to deny its existence by arguing that the trade-off is a subjective manifestation rather than an independently existing constraint \citep{cooper_emergent_2021,selbst_fairness_2019,friedler_impossibility_2016,friedler_impossibility_2021}. For instance, authors in \citep{cooper_emergent_2021} argue that the conflict between ``accuracy" and ``fairness" is the result of framing the trade-off as an optimization problem. Their argument rests on a causal fallacy. Recognizing the ``inherent conflict" between statically accurate outcomes and group similar outcomes in a data setting which contains group disparities and then optimizing between those competing interests cannot be the \emph{cause} of differences between subgroups of a target population that exist independently in that data setting. The realization that there are group disparities in a target population is the realization that there are differences between groups. The question of what causes differences between groups is beyond the scope of this article. However, to whatever extent subgroups of a target population are different, there will be differences in group outcomes no matter how they are measured. Changing the framing or values that guide an automated procedure (for instance by changing the definition or measurement of economic success) would simply lead to a different group skew in outcomes. Disparities in outcomes are inherent external constraints on any categorization or \emph{classification} of a world filled with group disparities.\footnote{Let us not forget that grouping based on sensitive attributes defined by the protected classes in non-discrimination law are not the only grouping that could be argued morally irrelevant.} The only way to remove disparities in outcomes between groups is to not recognize the differences between groups in the target population, and thus between individuals as well (i.e. to alter the data sample or its processing to reflect a false world.)

It is true that the trade-off is most commonly formulated as an optimization problem. For example, the lower bound of this trade-off has been estimated via proof \citep{gouic_projection_2020, zhao_inherent_2022}. And Authors in \citep{menon_cost_2018} have proven that in the case of a binary classifier it will be asymptotically possible to maximize both accuracy and fairness simultaneously only if the sensitive attribute and the target label are perfectly independent. On the other extreme, if the sensitive attribute is highly correlated with the target variable then it's only possible to maximize either the accuracy or the fairness at the same time. In between those two extremes, the trade off is determined by the strength of the correlation between the target and the sensitive attribute. As the \citep{menon_cost_2018} proof states, if the sensitive attributes and the target variable are perfectly independent of one another, the more generalizable the model is, the more group parity will be present. 

Some authors use this fact to argue that accuracy and fairness are complimentary \citep{cooper_emergent_2021,dutta_is_2020, hellman_measuring_2019}; even going so far as to state that the ``fairness-accuracy trade-off formulation also forecloses the very reasonable possibility that accuracy is generally in accord with fairness" \citep[p.4]{cooper_emergent_2021}. Statistically accurate outcomes and group similar outcomes are not generally in accord. While it is true that under certain conditions statistically accurate outcomes and group similar outcomes are complimentary, the reliance on that truth to minimize the importance of the trade-off is highly misleading. Statistically accurate outcomes and group similar outcomes can \emph{only} be complimentary in a data setting which is strictly group similar (necessarily defined as group parity in the context of perfect independence). If the data setting is already strictly group similar, there is no need for the fair machine learning approach. Fair machine learning is only required in those instances where unacceptable group dissimilarity exists, and in those instances, statistically accurate outcomes and group similar outcomes will always be uncomplimentary (i.e. the sensitive attributes and target variable will be correlated.) 

Others observe that, in practice, constraining outcomes to meet an acceptable notion of group dissimilarity can sometimes increase accuracy \citep{wick_unlocking_2019}. Again, the observation is correct but can lead to misunderstandings. When the use of a fairness constraint increases the accuracy, either the sensitive feature and target variable are independent (and so see the above argument) or the data sample was so unrepresentative that enforcing group similarity increased the accuracy by happenstance. And, that increase in accuracy by happenstance could never go beyond the group similarity present in the target population without decreasing the generalizability of the model.

Authors in \citep{dutta_is_2020} suggest that horizontal data collection can alleviate the trade-off.\footnote{Horizontal data collection is the collection of more features.} There, the authors are falling into the trap of thinking that group dissimilarity in outcomes is the result of inaccuracies; in this case,
an incomplete feature selection process where all the relevant-for-the-task features are not collected. Other researchers suggest that the trade-off can be alleviated by the targeted, vertical collection of data, where the model is trained on an enlarged dataset that serves the objective of group parity \citep{chen_why_2018, bakker_fairness_2019}. This will indeed improve group similarity. However, it will also result in a sample of data that is far from being representative of the real world distribution. Moreover, calculating the accuracy on an unrepresentative test set is misleading since the generalizability of the model when deployed may be compromised by this practice. In other words, these practices curate a data sample that satisfies group similarity, and then test the model on a test set which has also been curated. Furthermore, if the purpose of collecting data is not to use the target population as an external constraint, but instead to satisfy a notion of acceptable group dissimilarity, then there is no need to waste time and money collecting more data\textemdash simply use the existing processing techniques listed in section \ref{fairnessmetrics} to achieve the same result. 

Throughout this section, we have highlighted fallacies, misleading assertions, and questionable practices that result from the failure to understand the relationship between statistically accurate outcomes and group similar outcomes, and we've explained that a failure to understand that relationship often leads researchers to insist that group dissimilarities must be the result of inaccuracies. In the following subsection, we will, to the best of our ability, faithfully describe one of the most highly influential examples of this phenomenon, the model of automated decision-making procedures presented by Friedler et al. in \citep{friedler_impossibility_2016,friedler_impossibility_2021}. We believe their view requires careful attention.

\subsection{Critiques of the Model of Friedler et al.}
\label{framework1}
The model of Friedler et al. is based on two distinctions \citep[p.138-9]{friedler_impossibility_2021}. First, they make a distinction between the \emph{feature space}, which contains information about people, and the \emph{decision space}, which contains the decisions made about people based on the feature space. In other words, a decision making procedure takes inputs from the feature space and returns outputs in the decision space. Second, they make a distinction between the \emph{construct space} and the \emph{observed space}. The construct space consists of idealized representations about people and the decisions regarding them; whereas the observed space contains only the kind of information that is observable and the decisions that are the result of those observations. These two distinctions allow for the introduction of four combination spaces they call: (1) the \emph{construct feature space} which contains feature constructs like intelligence or frugality (CFS); (2) the \emph{observed feature space} which contains the observable or measurable features that are used as proxies for the construct features like IQ or debt-to-credit ratio (OFS); (3) the \emph{construct decision space} which contains decision constructs like college success or creditworthiness (CDS); and (4) the \emph{observed decision space} which contains decision proxies like college GPA or loan default (ODS). 

The observed spaces are knowable, and the construct spaces are unknowable \citep[p.3-6]{friedler_impossibility_2016}. More specifically, the distances between individuals and groups in the observable spaces are knowable and in the construct spaces unknowable. According to Friedler et al., two worldviews emerge from the picture once construct spaces are introduced: (1) the What You See Is What You Yet (WYSIWYG) worldview and (2) the Structural Bias worldview. The WYSIWYG worldview assumes that the observed features and decisions are essentially the same as the construct features and decisions; or, in other words, that the observations approximate the constructs (loan default approximates creditworthiness). Whereas, the Structural Bias (SB) worldview wishes to \emph{find} and define ``structural bias'' and ``non-discrimination'' in the transformations between these two spaces. More specifically, they wish to find group skew, defined as more distortion between groups than there is within groups in the transformations between an ``unobservable'' space and an observable space \citep[p.7]{friedler_impossibility_2016}. ``[T]o address this \emph{complication} . . . assumptions must be introduced about the points in the construct space, or the mapping between the construct space and observed space, or both'' [emphasis added] \citep[p.6]{friedler_impossibility_2016}. ``Structural bias'' is defined as group skew between the unknowable construct and knowable observed spaces in general \citep[p.7]{friedler_impossibility_2016}, and ``non-discrimination'' is specifically defined as an acceptable amount of group skew between the unknowable construct spaces and the knowable observed decision space \citep[p.8]{friedler_impossibility_2016}. 

Set aside, for a moment, these unknowable construct spaces and focus on their definition of ``direct discrimination.''  Under the SB worldview, direct discrimination is defined as the existence of group skew between the knowable, observed feature and decision spaces \citep[p.8]{friedler_impossibility_2016}, which means that direct discrimination can only be the result of a model that produces outcomes that are not representative of its sample. In other words, direct discrimination is the result of statistically inaccurate outcomes rather than the result of group disparities present in a representative sample (i.e. a problem the traditional machine learning approach can address).
Note that in all the definitions listed above, group skew is the quotient of ``between group distance'' and ``within group distance''. In other words, to say that there exists group skew is to say that groups are different.

Again, structural bias and non-discrimination are the result of group skew in the transformations between the construct and observed spaces. Unlike direct discrimination, group skew in both of these cases is based on unknowable, assumed mappings. It is in this way, according to Friedler et al., that observational processes of an automated decision procedure cause group skew \citep[p.140]{friedler_impossibility_2021}. We wish to temper their assertion: if (1) the assumption that a construct space must exist and (2) unknowable knowledge about that space is granted, then the conclusion that observational processes cause group skew can be reached. At first glance, it may appear that the statement that observational processes cause group skew is supporting the ``framing'' argument rejected previously; however, the argument there was that the framing of the relationship between accuracy and fairness as an optimization problem is the cause of the trade-off between the two. 

There is also a difference between: (1) \emph{group skew} that is the result of more distortion between groups than there is within groups in the transformations between construct and observed spaces and (2) \emph{group dissimilarity} that is the result of group disparities in a representative sample. In other words, group skew that results from observational approximations of constructs does not cause group dissimilarity in outcomes \emph{generally} but only ever  \emph{additionally} in a data setting that contains group disparities (i.e., those settings where fair machine learning is required). Group skew defined in the SB worldview is addressed, not by producing group similar outcomes in an ``unfair'' data setting, but instead by selecting observations which accurately (or ``correctly'') approximate constructs. However, the approximation is subject to unknowable information and rests on the assumption that decision-makers only ever want to satisfy abstract standards like creditworthiness or academic success rather than specific standards like loan default or college GPA. It is in no way clear to us that the introduction of unknowable construct spaces with definitions that mimic legal terminology produce anything more than confusion.

Regardless, the definitions of structural bias and non-discrimination, developed by Friedler et al., are irrelevant for the equitable, fair machine learning approach. Here is yet another instance of researchers falling into the trap of thinking that group dissimilarity has to be the result of inaccuracies, this time ethereal inaccuracies. As stated previously, the goal of fair machine learning approach is not to produce statistically accurate outcomes nor ethereally ``correct'' outcomes, but instead to produce equitable, group similar outcomes in data settings that would otherwise produce inequities. In reality, their model does not address issues of equity in automated decision-making systems.

\section{Evaluation of the Trade-off}\label{framework2}
This section provides a numerical proof-of-concept evaluation of the trade-off between statistically accurate outcomes and group similar outcomes. First, the used data samples and pre-processing techniques will be described and then the results will be discussed.

\subsection{Experimental Setup} 
In order to study this trade-off, we synthetically generate three datasets with  group disparities where the probability to receive a favourable outcome is not equal for all groups. Each dataset is composed of 3 variables: group (G) representing a dependant and sensitive, binary attribute, value (V) the insensitive variable and outcome (Y)  as a binary label or targeted value which has two values 0 and 1 for the unfavourable and favourable outcomes respectively.
Table \ref{datasets} describes the characteristics of each dataset. For fairness metrics, we use three group fairness metrics, namely: disparate impact, equalized odds, and statistical parity. In our experiment, we used pre-processing techniques to remove group skew, and we show how these techniques affect the statistical accuracy of outcomes in a data sample where the skew is large. We can clearly see that the resulting distributions will be very different than the original distribution of data. Group skew is computed as the ratio between the between-groups variance and the within-groups variance:
\[ GS = \frac{\sigma_{B}^2}{\sigma_{W}^2}\]
where:
\[\sigma_{B}^2 = \sum_{i=1}^{N_g}  N_{i} * (\mu_i - \Bar{X)}\]
\[\sigma_{W}^2 = \sum_{j=1}^{n}  (x_{ij} -\mu_i) \]
Where GS denotes group skew, $\sigma_{B}^2$ is the between group variance and $\sigma_{W}^2$ is the in-between group variance. $N_g$ is the total number of groups and $n$ is the number of observations in each group $i$.
Between group Variation $\sigma_{B}^2$ is the total variation between each group  mean $\mu_i$ and the overall mean $\Bar{X}$.
Within-group variation is the total variation in the individual values in each group $x_{ij}$and their group mean$\mu_i$.

\def\arraystretch{1.1}

\begin{table*}[h] 
 
    \begin{tabularx}{\textwidth}{|X | X| X | X|  X  | X | X|  X  | X ||} 
 \hline
 Dataset & group & size & mean & std &Pr(Y= 1)& statisti- cal parity & Dispar- ate impact & group skew\\ [0.5ex] 
 \hline\hline
 \multirow{2}{*}{D1} & 1 & 10000& 5.5 & 0.6 & 0.7 & -0.20 & 0.72 & 2356.41\\ 
 \cmidrule(l){2-6}
  & 0 & 10000  & 6  & 0.4 & 0.5 &  & & \\
 \hline
 \multirow{2}{*}{D2} & 1 & 20000 & 5.5  & 0.3 & 0.8 & -0.59   & 0.25  & 2332.88\\ 
 \cmidrule(l){2-6}
  & 0 & 9000 & 6 & 0.6 &  0.2 &  &  &\\
 \hline
 \multirow{2}{*}{D3} & 1 & 10000 & 5.5 & 0.6 & 0.5  & 0.01  & 1.02 & 2790.86 \\ 
 \cmidrule(lr){2-6}
  & 0 &10000 & 6 & 0.3 & 0.5  &  & &\\
 \hline
 
\end{tabularx}
\caption{Overall overview about the used datasets}
\label{datasets}
\end{table*}

\textbf{The Disparate Impact Remover:}
to remove disparate impact(we note DIR(repair value)) we use the pre-processing algorithm proposed by Feldman et al. \citep{Feldman_disparate_impact} with 3 different repair levels 0.3 , 0.5 and 1.0 to control the degree of overlap of the distributions of the two groups, results in table \ref{tab:Results disparate impact}.

\textbf{The Equalised Odds Remover:} to remove the effect of equalised odds we experimented using 3 different algorithms namely: (1) reweighing algorithm \citep{kamiran2012data}, (2) Fair Balance \citep{Yu2021FairBalanceIM} and FairBalanceVariant \citep{Yu2021FairBalanceIM}, results in table \ref{tab:Results equalised odds }.

\textbf{The Statistical Disparity Remover:} in order to remove the statistical disparity from the data, we used the algorithm proposed by Zemel et al. \citep{zemel2013learning}, results can be found in table \ref{tab:Results stat parity}.

\begin{table*}[]
\begin{tabularx}{\textwidth}{|X|X|X|X|} 
 \hline
 Dataset & DIR (0.3) & DIR (0.5) & DIR (1.0)  \\ [0.5ex] 
    \hline
    \hline
    D1 & 1268.29 & 710.91 & 2.30 \\
    \hline
    D2 & 1257.20 & 705.24  & 3.51 \\
    \hline
    D3 & 1632.39 & 958.50& 4.73 \\
    \hline
    \end{tabularx}
    \caption{Group skew results when using Disparate Impact Remover (DIR)}
\label{tab:Results disparate impact} 
\end{table*}
\begin{figure}[H]
\includegraphics[width=\textwidth]{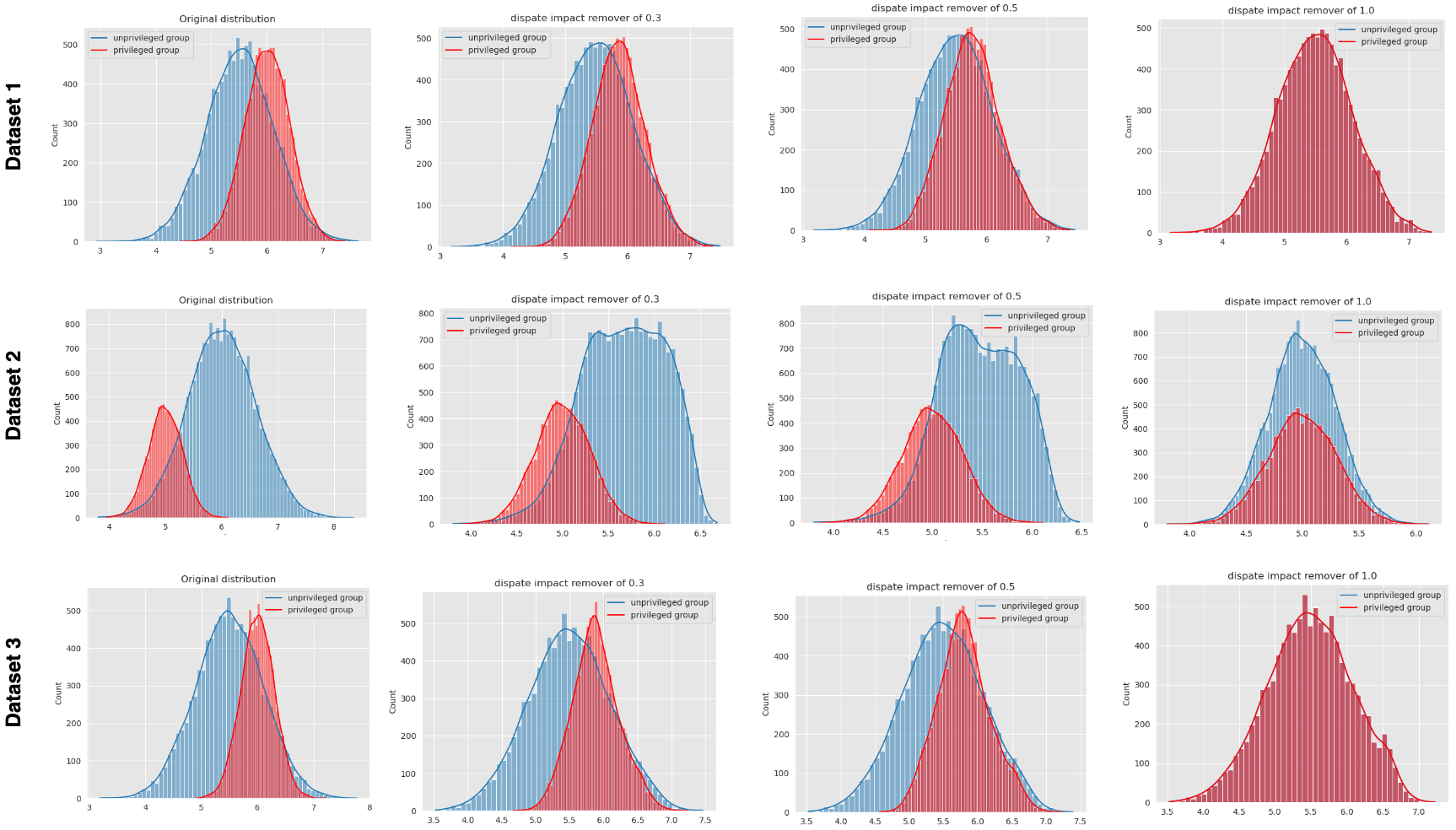}
\caption{Histograms of the different datasets and their transformed versions obtained when running Disparate Impact Remover \citep{Feldman_disparate_impact} with repair values of 0.3 , 0.5 and 1.0}
\label{fig:DI}
\end{figure}

\begin{table*}[t]
\begin{tabularx}{\textwidth}{|X|X|X|X|} 
 \hline
 Dataset & Reweighing algorithm & FairBalance algorithm  & FairBalanceVariant algorithm  \\ [0.5ex] 
    \hline
    \hline
    D1 & 369.52 & 167.19 & 167.19 \\
    \hline
    D2 & 325.51 & 144.88 & 144.88\\
    \hline
    D3 & 2720.71 & 2717.75 & 2717.75\\
    \hline
    \end{tabularx}
    \caption{Group skew results when using Equalised Odds removers}
\label{tab:Results equalised odds } 
\end{table*}
\begin{figure}[H]
\includegraphics[width=\textwidth]{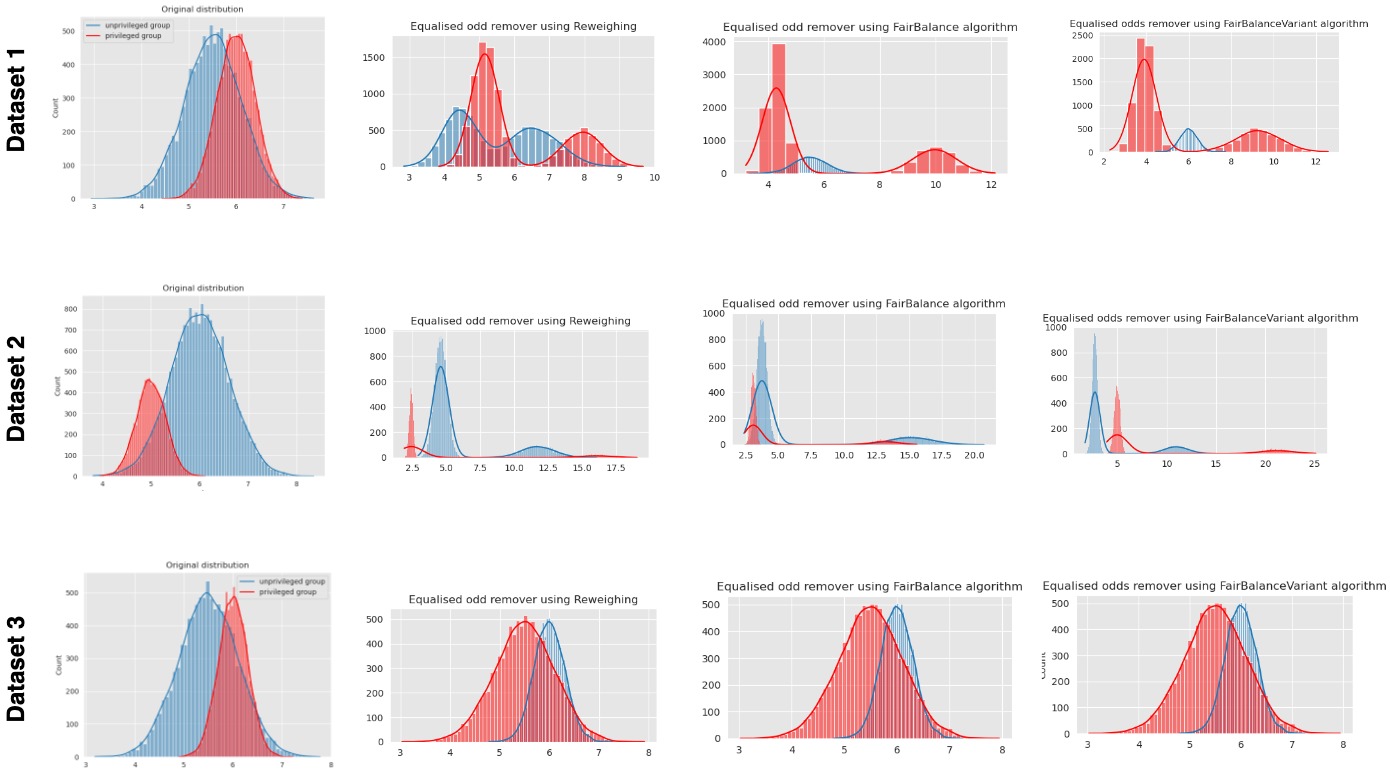}
\caption{Histograms of the different datasets and their transformed versions obtained when running the different Equalised Odds removers ((1) reweighing algorithm \citep{kamiran2012data}, (2) Fair Balance \citep{Yu2021FairBalanceIM} and FairBalanceVariant \citep{Yu2021FairBalanceIM}) }
\label{fig:EODDsDist}
\end{figure}

\begin{table*}[]
\begin{tabularx}{\textwidth}{|X|X|} 
 \hline
 Dataset & Statistical Disparity Remover \\ [0.5ex] 
    \hline
    \hline
    D1 &  693.32\\
    \hline
    D2 &  693.32\\
    \hline
    D3 & 1020.47\\
    \hline
    \end{tabularx}
    \caption{Group skew results when using Statistical Disparity Remover}
\label{tab:Results stat parity} 
\end{table*}

\begin{figure}[H]
\includegraphics[width=\textwidth]{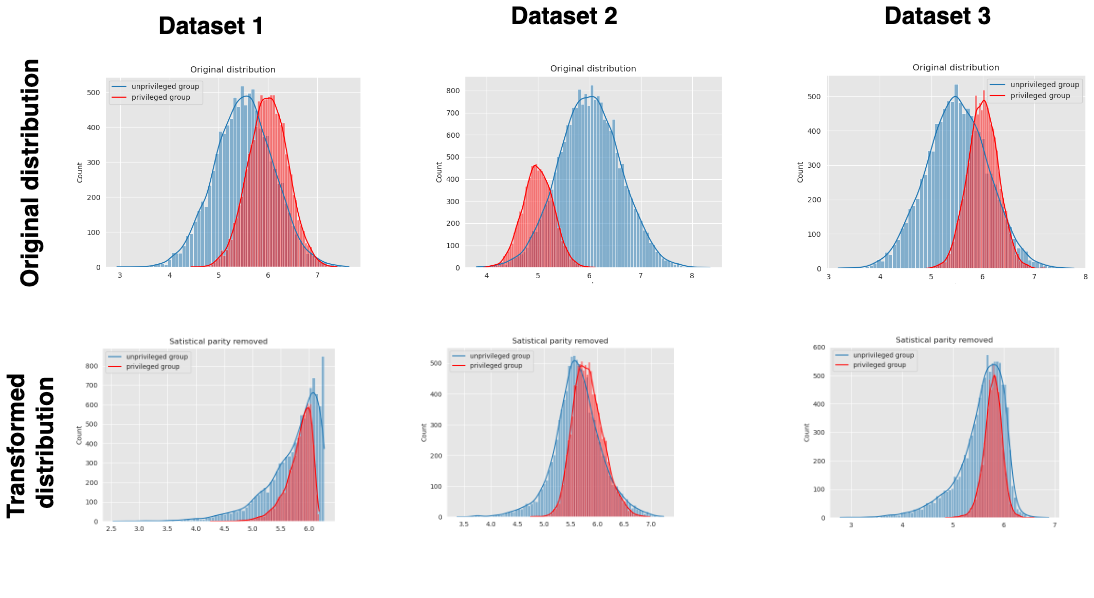}
\caption{Histograms of the different datasets and their transformed versions obtained when running the Statistical Parity Remover \citep{zemel2013learning}.}
\label{fig:statParity}
\end{figure}

\subsection{Discussion}
The Disparate Impact Remover (see table \ref{tab:Results disparate impact}) with a repair scale of 1.0 achieved the lowest group skew for the three data samples by ensuring a complete overlap between the groups. The positive point about these techniques is that they preserve the ranking between groups which ensures less distortion in the transformed versions of the data (see figure \ref{fig:DI}). However, this technique results in the highest group skew differences between the transformed sample and the original which suggests that the outcomes will not be statistically accurate (see section \ref{tradeoff}), assuming that the original sample is representative. 

For the equalised odds removers, we find that the FairBalance algorithm and its variance \citep{Yu2021FairBalanceIM} achieved a low group skew on the three datasets. Specifically for this set of techniques, we observe that they change drastically the shape of the resulting distribution (see figure \ref{fig:EODDsDist}) except for dataset 3. For D3, the resulting transformed D3 is less distorted compared to the transformed versions of D1 or D2. D3 contained an initial group skew but the sample was perfectly balanced since the size of each group is equal and the probability of receiving a favourable outcome was equal for each group. This is an indication that the equalised odds removers are sensitive towards group imbalances. The more imbalanced the groups, the bigger the distortion in the transformed dataset. 

The Statistical Disparity Remover \citep{zemel2013learning} scored the worst among all techniques in decreasing the group skew except for the data sample 3, D3 where it scored better than all the equalised odds removers.

The  distortions that can be observed in the transformed samples when using pre-processing techniques in Figures \ref{fig:DI}, \ref{fig:EODDsDist} and \ref{fig:statParity} show an illustration of how much can a fairness guided learning process reflect a different reality, models receiving and tested on those transformed samples may suffer from a lack of generalizability where the test results are not reliable.

\section{Conclusion}

First, we hope to have shown that the fair machine learning community represents and advocates for a single conception of fairness, defined as equity in automated decisions. We suspect that the unipolar conception of fairness in the fair machine learning literature is due to the fact that the traditional machine learning approach has the potential to satisfy a meritocratic conception but can never satisfy an equitable conception in a data setting that contains group disparities. Our concern is that by only defining fairness as equity in machine learning, the community may be leading policy-makers and regulators to believe that \emph{fairness} is absent in automated decisions without the use of the equitable, fair machine learning approach. Note that, throughout this article, we have neither advocated for nor disparaged any particular conception. We hope for an expansion in how fairness is conceived in machine learning, so that the literature can capture the same kind of diversity in opinion that is present in the wider societal discourse.

Second, we hope to have shown that the rejection of the trade-off between statistically accurate outcomes and group similar outcomes as an independent, external constraint has resulted in fallacious reasoning, misleading assertions, and/or questionable practices. Researchers should confront the reality that equitable outcomes require the introduction of inaccuracies. Admitting that there exists a trade-off does not mean that outcomes should not be equitable. We argue, though, that the research community should be more straightforward about what is being sacrificed in the name of equity. Obfuscating the nature of that sacrifice, for instance by redefining the term ``bias" in machine learning, could be misleading policy-makers and regulators. As was once wisely said, ``There are no solutions. There are only trade-offs" \citep{sowell_conflict_1987}.

To those ends and in an effort to foster transparency, we introduced experimental results to aid designers of automated decision-making systems in understanding the relationship between statistically accurate outcomes and group similar outcomes. Future work could use the conceptual and experimental understanding provided throughout this article in a variety of relevant disciplines: (1) data scientists might build a toolkit that would allow researchers and designers of automated decision procedures to incorporate goals and compromises into the machine learning pipeline, where primary and secondary goals are represented by chosen fairness metrics and distributions; (2) legal scholars might realize that, perhaps, affirmative action and positive action are the applicable legal doctrines rather than disparate impact and indirect discrimination, because fair machine learning techniques alter (act on), the decision process itself (i.e., the implementation of fair machine learning metrics is positive discrimination and so it must be determined whether that positive discrimination constitutes lawful or unlawful discrimination in a given jurisdiction and decision context); and (3) data ethicists might offer an alternative proxy for (un)fairness in the machine learning pipeline, other than group similarity (skew as the quotient of between-group and in-group distances), that could allow for a conception of fairness that is not based on equity.

\bmhead{Acknowledgments}

The research presented in this paper has received funding from the European Union's funded project LeADS under Grant Agreement no. 956562. We would like to give special thanks to Gabriele Lenzini, Jean-Michel Loubes, and Maciej Zuziak for their advice and feedback throughout the writing process.

\bibliography{sn-bibliography}

\end{document}